\documentclass[a4paper]{article}

\usepackage{icrc2013}

\title{The VERITAS Upgraded Telescope-Level Trigger Systems: Technical Details and Performance Characterization}

\shorttitle{VERITAS Telescope-Level Trigger System}

\authors{
Benjamin Zitzer$^{1}$,
for the VERITAS Collaboration.
}

\afiliations{
$^1$ Argonne National Laboratory \\

}

\email{bzitzer@anl.gov}

\abstract{VERITAS is an array of imaging atmospheric Cherenkov telescopes sensitive to gamma rays in the energy range between $\sim$100 GeV and $\sim$50 TeV. The instrument underwent an upgrade of the camera triggers in November 2011. The new systems use 400 MHz Xilinix Virtex-5 FPGAs for the pixel neighbor coincidence logic necessary to produce a camera-level trigger. The upgraded systems are capable of time-aligning individual triggering pixels to within $\sim$0.2 nanoseconds, allowing for an operational pixel-to-pixel coincidence window of $\sim$5 nanoseconds. This reduced coincidence window provides improved rejection of night-sky background (NSB) which permits a reduction of the energy threshold at the trigger level. The use of FPGAs allows for the future implementation of a topological trigger capable of discriminating events based on an image moment analysis of a bit-wise hit pattern. As part of the commissioning phase for the trigger upgrade, the hardware was initially installed in a single telescope in "parallel" to the (then) current system. This allowed for the detailed performance characterization of the new system relative to the pre-existing trigger. Here we present technical details of the upgraded VERITAS camera trigger system and outline the details of these performance studies.}


\begin{document}
\maketitle

\section{Introduction}
VERITAS (Very Energetic Radiation Imaging Telescope Array System) is an array of four imaging atmospheric Cherenkov telescopes (IACTs)  located in southern Arizona, USA, for observing the northern sky in very-high-energy (VHE) gamma rays (above 100 GeV). IACTs detect the Cherenkov light emitted from particles interacting in the Earth’'s atmosphere. Primary particles (such as gamma rays and cosmic hadrons) interact in the upper atmosphere, creating showers of secondary particles. Energetic particles in the shower give rise to a Cherenkov light pool that can be seen by IACTs such as VERITAS. Each VERITAS telescope employs a 12m diameter tessellated mirror to reflect the Cherenkov images onto cameras composed of 499 Photomulitplier Tubes (PMTs) \cite{Holder2006}. 

IACTs are limited at the lowest energies because of the steeply rising trigger rate produced by the night-sky background (NSB) light (starlight) and by cosmic rays (CRs, dominated by protons, muons and electrons).  Cherenkov air showers that are initiated by gamma rays are typically seen as roughly elliptical in a single telescope with a duration of less than 10 ns. NSB events occur in single PMTs with random timing with respect to other NSB events. Showers initiated by CR protons are typically larger and rounder than gamma-ray events. Muon events typically appear as arcs or rings in the camera. VERITAS employs a three-level trigger system to reduce the number of background events.  At the first level of triggering (L1), a constant-fraction discriminator (CFD) requires a PMT pulse height above a programmable threshold (typically set around 5-6 photoelectrons). The second level of triggering (L2, also called the pattern trigger or the telescope-level trigger) requires a L1 signal in at least three adjacent PMTs within a timing coincidence window. A third level of triggering (L3) requires a L2 trigger in at least 2 telescopes within a 50 ns coincidence window. Events which pass the L3 trigger are readout by the data acquisition system and recorded for use in the offline analysis. The data acquisition consists of 500 Msample/s FADCs, which digitize the PMT waveforms and stores them in a 64$\mu$s memory buffer \cite{Holder2006}. The telescope-level trigger for VERITAS was replaced in November 2011 with a FPGA-based system.

At the single-telescope level, exploiting the event topology for gamma-ray events is not new. The Whipple 10m and the University of Durham telescopes, which were a pioneering instruments in the field of IACTs, employed a nearest-neighbor logic trigger requirement before a similar design was implemented for VERITAS \cite{Holder2006},\cite{Leeds2002},\cite{Durham1999}. The upgraded telescope-level trigger, with more modern technology, including fast FPGAs, provides better pixel-to-pixel timing alignment and allow for a narrower coincidence time width compared to the then-current VERITAS trigger, but still requires the 3-fold neighboring pixel requirement.

\section{Design of Trigger System}

A photograph of the L2 crate for the VERITAS upgrade is shown in Figure 1. Each 9Ux160mm L2 crate contains three types of boards: ten input boards, three L1.5 boards, and one L2 board. A block diagram of the L2 is shown in Figure 2. The design of this trigger system has been discussed before in \cite{JTA2008}, \cite{Krennrich2008}, \cite{ZitzerTipp2011}. 

\begin{figure}[h]
	\centering
	\includegraphics[width=3.2in]{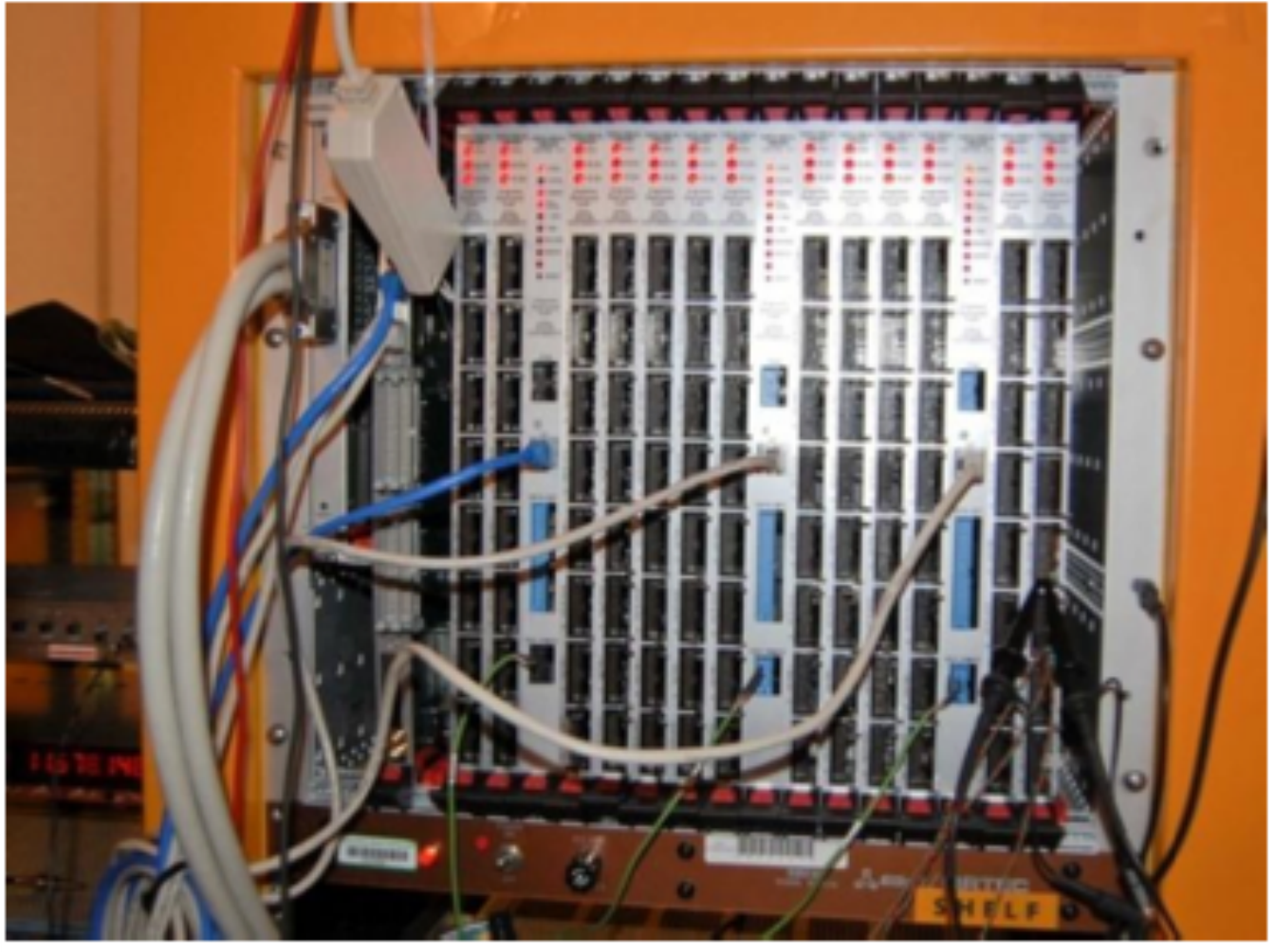}
	\caption{A photograph of the L2 crate for the VERITAS upgrade. The signal cables from the CFDs are unplugged.}
	\label{L2Photo}
\end{figure}

\subsection{Input Boards and Backplane}

Each of the 499 L1 outputs is a 13 ns wide emitter-coupled logic (ECL) pulse . These signals are routed to the input boards, which translate the ECL signals to Low-voltage differential signaling (LVDS). The signals are then sent through the high-speed custom VME backplane to the three L1.5 boards. The camera is divided into three L1.5 regions as shown in Figure 3. Pixel signals in overlapping L1.5 regions are copied to both neighboring L1.5 boards to give a relatively flat triggering efficiency over the entire camera. 

\begin{figure}[h]
	\centering
	\includegraphics[width=0.5\textwidth]{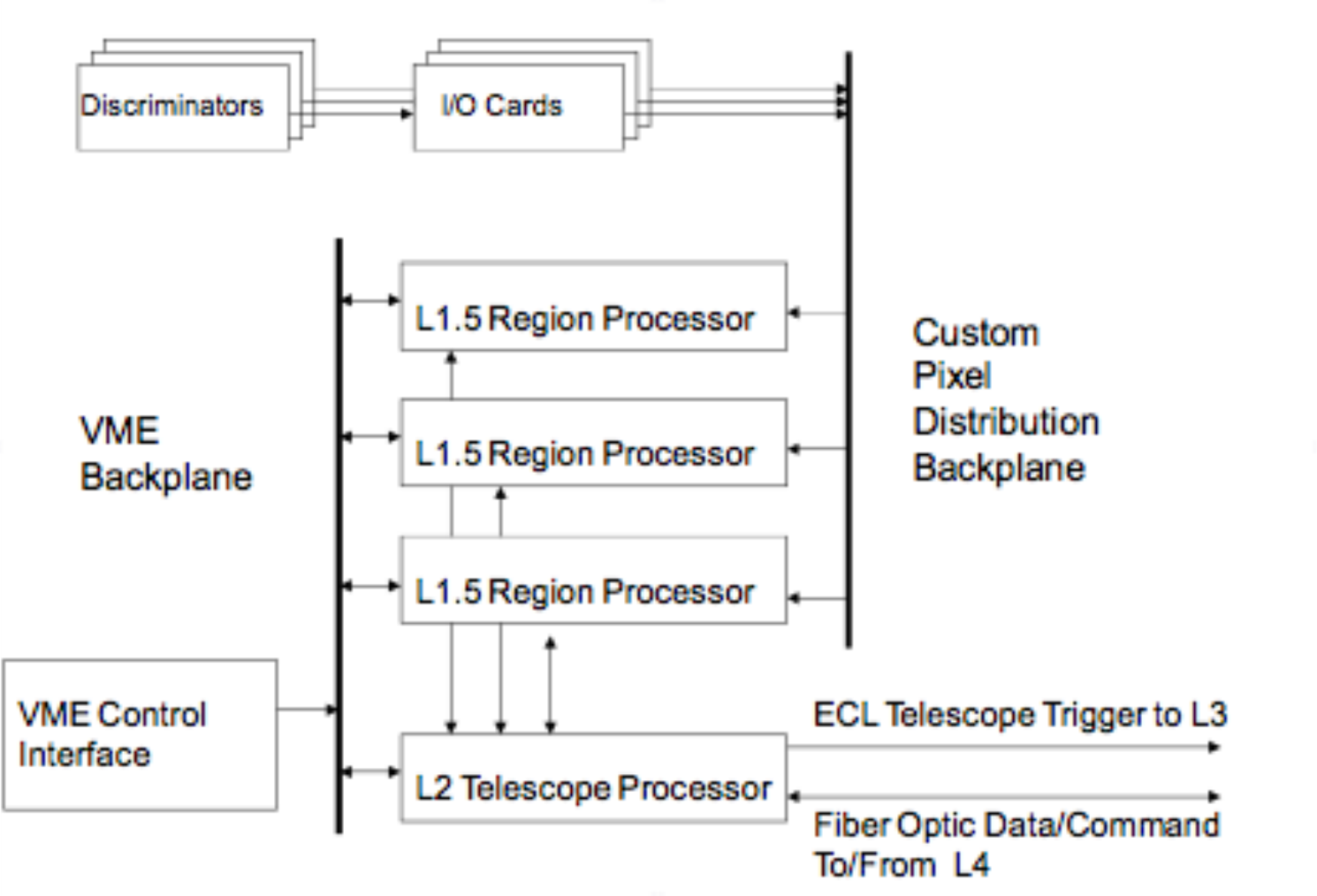}
	\caption{Block diagram of the pattern trigger for the VERITAS upgrade. Note that the fiber optic data/command to/from L4 is not currently in use, but could potentially be in used for a future VERITAS upgrade.}
	\label{L2Block}
\end{figure}

\subsection{L15 Boards}

The L1.5 board contains a Xilinx Virtex-5 FPGA that processes the coincidence neighbor logic. The Virtex-5 was chosen for its higher speed and specific cell structure. The higher speed allows for the logic to run significantly faster than the original system. The specific cell structure of the Virtex-5 allows computation of the coincidence equations in individual cells, resulting in a significantly enhanced ability to control the delay of each pixel before entering the coincidence equation. The usefulness of controlling the delay is discussed in the pixel timing alignment section.

Each L1 signal is a the center of a cell within the FPGA consisting of itself and up to six neighboring pixels around it. An L1 signal of the center pixel in the cell and two of the neighbor pixels is required for the trigger. The L1 signal for the pixels located in a overlap region (see Figure 3) may have two or three cells associated with it. The trigger is asynchronous, requiring only minimal overlap time between neighboring pixels before a trigger occurs, meaning that at no point is the data sampled. An additional programmable required overlap time called ``detune" is used to control the coincidence gate width. The speed of the trigger is therefore only limited by the propagation delay within the FPGA and the lookup speed of the memory.  Trigger bits that pass the neighboring pixel and timing requirements are sent to the L2 processor.  

\begin{figure}[h]
	\centering
	\includegraphics[width=3in]{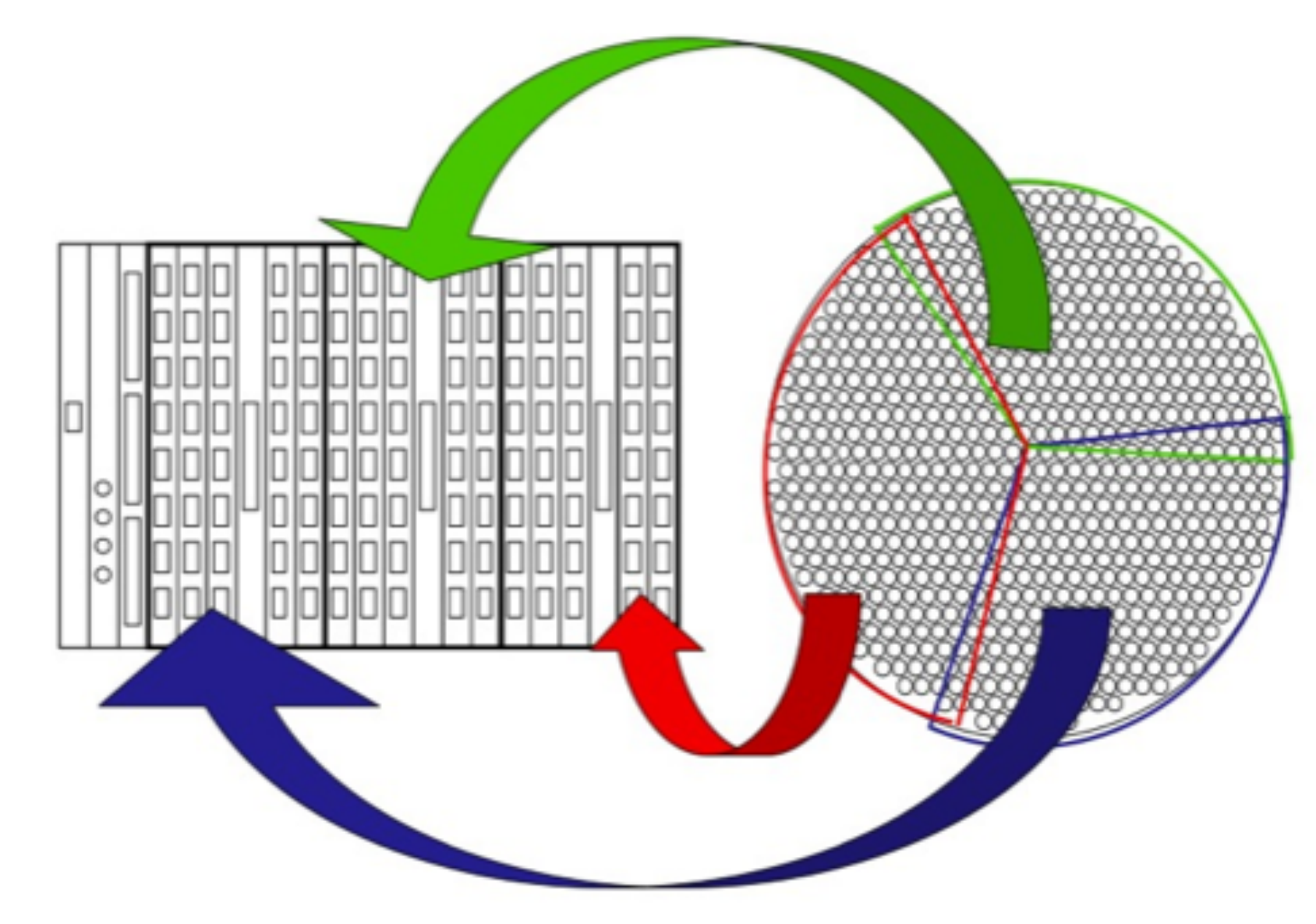}
	\caption{Illustration showing the division of a VERITAS camera into three L1.5 regions.}	
	\label{L15Division}
\end{figure}

\subsection{L2 Board}

The L2 board in each telescope-level trigger crate serves as an OR gate between L1.5 boards, sending the L1.5 trigger bit to the L3 array trigger. The Virtex-4 FPGA in the L2 board also contains two time-to-digital converters (TDCs), each with time resolution of $\sim$50ps, which are utilized in the timing alignment procedure described in the timing alignment section. It provides the clock source for the L1.5 boards and has the capability of calculating the image moments ($n$, $\Sigma x$, $\Sigma y$, $\Sigma x^{2}$, $\Sigma y^{2}$, $\Sigma xy$) required for the topological trigger described briefly in the conclusions section. 

\section{Trigger System Performance}

The telescope-level trigger for VERITAS was replaced in two days in November 2011, during full moonlight when the telescopes were not in operation, so there was no loss of science data to the experiment during installation. The performance and added benefits of the new trigger system are described in this section.

\subsection{Efficiency Studies}

Prior to the installation in November 2011, one of the new telescope-level trigger crates was installed in one of the telescopes to monitor the trigger efficiency \emph{in situ} across the camera. The CFD cables were connected to the new telescope trigger and ECL signals were copied to the primary (pre-upgrade) primary telescope-level system using modified input boards (referred to as I/O boards). The trigger bit output (L2 signal) of the new telescope-level trigger was then sent to the FADCs into a channel with a dead PMT. This allowed event-by-event comparisons of the two triggers in various operating conditions, with a minimal amount of time lost to the experiment. Event topology and relative efficiency of both systems could then be explored in the offline analysis.

\begin{figure}[h]
	\centering
	\includegraphics[width=3.2in]{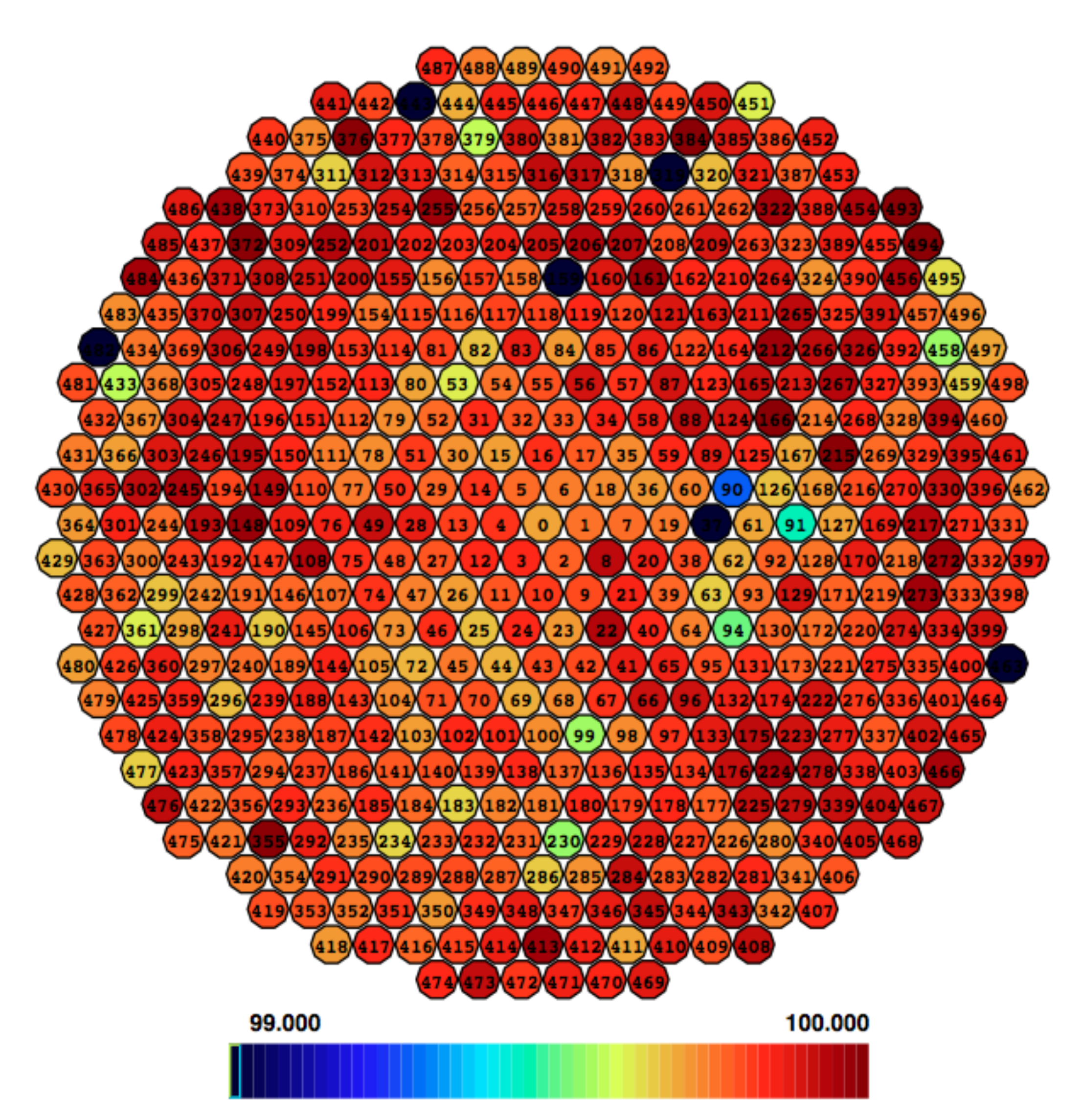}
	\caption{Efficiency over a VERITAS camera map of the upgraded telescope-level trigger to the pre-upgrade trigger. Each pixel shows the ratio of number of CFD hits of both triggers firing to only the pre-upgrade trigger firing. Note that the scale is from 99\% to 100\% and that the ratio cannot exceed 100\%. Dead PMTs in the camera appear black in this map. }	
	\label{Eff}
\end{figure}

An example of one of the results of the efficiency study is shown in Figure 4. It shows the ratio of the L1 rates when both systems triggered on the same event to number of times the pre-upgrade system triggered. This efficiency ratio therefore cannot be over 100\%. Figure 4 shows that both systems trigger together for $\sim$99.5\% of all events used in the offline analysis. The coincidence width that was chosen for the data shown here was one where the coincidence widths of both triggers were closely matched ($\sim$9 ns). Figure 4 indicates no regions of inefficiency across the camera. This was typical for all VERITAS operations when both triggers were functioning properly and the upgrade trigger was time aligned (next section).

\subsection{Pixel Timing Alignment}

Without any sort of pixel alignment, the relative pixel-to-pixel difference in arrival time of leading CFD edges in the L1.5 board is within a window of a few nanoseconds, due to transit time differences of the PMTs, signal cables, CFDs, CFD cables, input boards and the routing in the backplane. In order to ensure that all CFD signals reach the coincidence logic within the L1.5 boards within a window of $\pm$1ns, a timing alignment procedure is implemented. This requires careful control over the skew and overall delay within the internal routing of the L1.5 boards, as well as the L1.5 FPGAs and the L2 FPGA coincidence signal. The L1.5 FPGAs can be programmed to delay individual input signals in steps of 72ps up to $\sim$10ns before they are sent through the coincidence logic. This careful control over the pixel-to-pixel timing allows to narrow the gate width to as little as $\sim$3ns while keeping the trigger efficency relatively flat over the camera's surface.

\begin{figure}[h]
	\centering
	\includegraphics[width=3.2in]{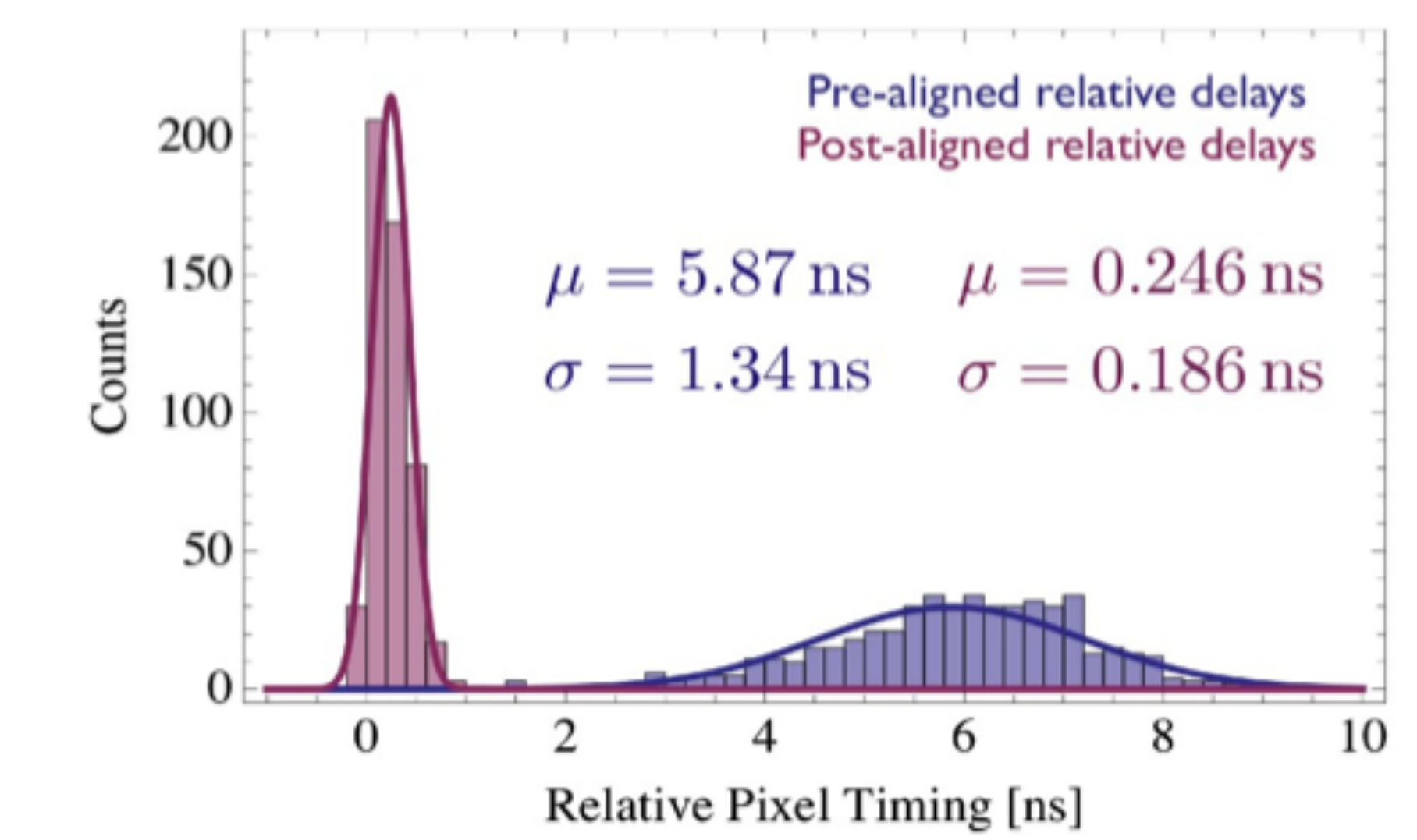}
	\caption{Histograms showing the relative delays of the CFD leading edges before and after the timing alignment procedure. The pixels are plotted relative to the slowest pixel arrival time in the distribution prior to timing alignment.}	
	\label{PixAlignment}
\end{figure}

The procedure for the timing alignment requires the LED flasher used for calibration running and the telescopes pointing at dark patch of sky at a high elevation to reduce the amount of NSB the telescopes are exposed to during the process. The CFD thresholds are set at a higher threshold to help reduce NSB contamination. Seven different 3-fold coincidence patterns per pixel are tested multiple times and averaged to find the mean arrival time for that pixel. The delay of that pixel is measured relative to the average of two fixed coincidence sets, one in each of the other two L1.5 regions of the camera. These fixed sets are a reference time for all coincidence sets. A delay time is then calculated and added to move that pixel arrival time closer to the mean of the reference coincidence sets.

Figure 5 shows the relative pixel-to-pixel timing for one of the telescopes before and after the alignment procedure. The gaussian $\sigma$ after the alignment is 186 ps, putting the level-2 for that telescope within the performance goal of $\pm$1ns. All four telescopes have a post-alignment Gaussian $\sigma$ of less than 200 ps. 

\subsection{Diagnostic Tools}

The upgraded telescope-level trigger has a VME control through a CORBA interface to a GUI on a PC, allowing the VERITAS observers to monitor and control the trigger during operation. The included features are: L1 rate monitors, enable/disable controls for each of the CFDs, control over the timing alignment and coincidence gate width. A pre-scaling factor also exists within the upgraded telescope-level trigger  that is also controlled by the GUI. The pre-scaling factor is used for taking single telescope runs dedicated to measuring CR muons. Due to their small collection area and spacing between the telescopes, requiring two telescopes for the array trigger during normal operations removes many of the muons at the trigger level.

\subsection{Adjustable Coincidence Width}

\begin{figure}
	\centering
	\includegraphics[width=3.2in]{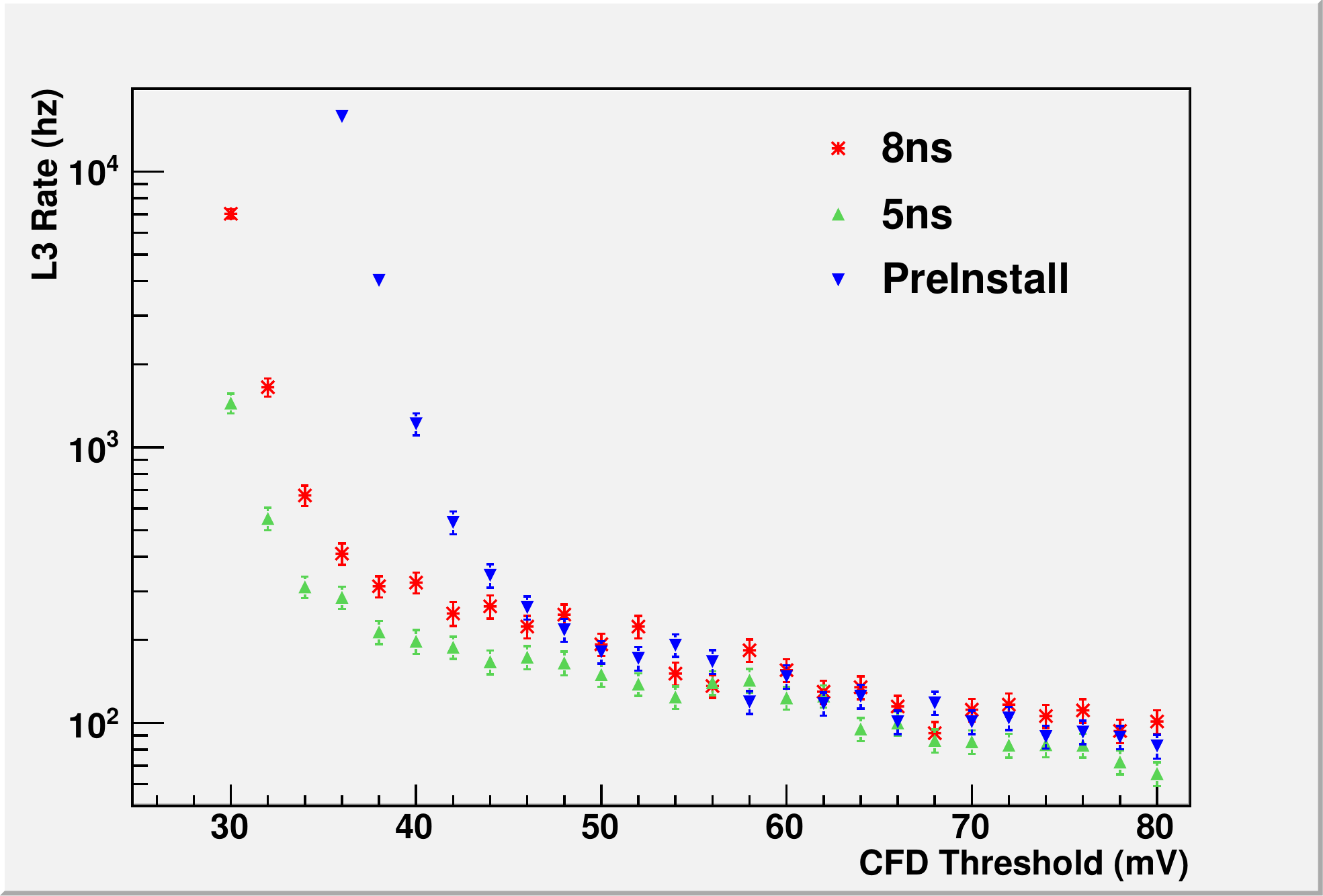}
	\caption{Bias curves of the array trigger rate plotted against the CFD threshold for the pre-upgrade level-2 system (in red) and then for the post-upgrade level-2 with a coincidence gate width of 8 ns (green) and 5 ns (blue). At lower CFD thresholds, the L3 rate is dominated by NSB, while at higher thresholds it is dominated by CRs.}	
	\label{L3Bias}
\end{figure}

The pre-upgrade telescope-level trigger has a fixed coincidence gate width of 8-10 ns, depending on telescope. This value was fixed and could not be changed after installation, along with the pixel-to-pixel timing. The new FPGA-based trigger includes firmware optimized for rapidly solving trigger equations and pixel alignment to $\pm$1 ns, and is capable of yielding a coincidence gate width down to 3 ns. The L1.5 FPGA has a programmable parameter called `detune' which is an extra required overlap width required for a trigger to occur. Increasing detune, therefore, has the effect of narrowing the maximum allowed coincidence gate width. Since NSB triggers occur at random times with respect to one another, narrowing the gate width has the effect of reducing the number of accidental NSB triggers. 

The effect of the narrower coincidence width on NSB accidental triggers is best illustrated by the bias curves shown in Figure 6. The bias curve is used by the VERITAS collaboration to determine the trigger scheme for the array. The L3 (or data acquisition) rate is plotted against the CFD threshold. NSB triggers dominate at the lower thresholds, while cosmic rays (including gamma rays) dominate at the higher thresholds. The difference between the 8 ns and 5 ns coincidence width bias curves show that the NSB rates are lower while keeping most of the CR events. The placement of the CFD threshold should be set right above the inflection point between the NSB-dominated and CR-dominated regimes. L3 rates that are higher than a few hundred hertz can significantly increase dead time and overwhelm the data acquisition system. Low energy gamma-ray events have the trend to be smaller and fainter than higher energy events, and are therefore more likely to be seen at lower CFD thresholds. Lowering the CFD thresholds has the overall effect of lowering the energy threshold of the instrument. Given that the NSB rates depend on the brightness of the sky, the optimal CFD threshold depends on different factors including the weather conditions, the intensity of star light in the FOV, and moonlight. The effect of the narrower gate width to lower the CFD threshold has been demonstrated in all of of these higher NSB conditions.  

\section{Conclusions}
The new telescope-level trigger has been successfully deployed at VERITAS as an upgrade to the existing hardware. In conjunction with the PMT camera upgrade \cite{ZitzerTipp2011} \cite{KiedaICRC2013}, VERITAS plans to lower the energy threshold of the instrument to $\sim$80 GeV. The upgraded trigger has demonstrated the ability to lower the energy threshold by pushing down the acceptable CFD threshold levels for all modes of operation, which was a major performance goal. The tighter pixel timing alignment allows for a narrower coincidence gate width down to 3 ns with uniform camera efficiency. Studies using Crab Nebula data with varying coincidence widths prior to the camera upgrade showed that 5 ns was optimal for gamma-ray efficiency. New diagnostic features allow VERITAS observers greater diagnostic and control that was not available before.

The upgraded trigger has the hardware capabilities for two potential VERITAS upgrades beyond this current one: a muon trigger and a topological trigger (or L4). The FGPAs in the L1.5 and L2 boards could be utilized to pick out simple ring or arc patterns in L1 signals that are strong muon indicators in IACT cameras. The L4 trigger requires that each L2 board would calculate the image moments and send that information to a central L4 processor which would discriminate hadron events from gamma-ray events at the hardware level. This has been proposed as the array-level trigger design for CTA and work is in progress to apply lessons learned from this trigger upgrade to CTA. 

\vspace*{0.5cm}
\section*{Acknowledgment}

This research is supported by grants from the U.S. Department of Energy Office of Science, the U.S. National Science Foundation and the Smithsonian Institution, by NSERC in Canada, by Science Foundation Ireland (SFI 10/RFP/AST2748) and by STFC in the U.K. We acknowledge the excellent work of the technical support staff at the Fred Lawrence Whipple Observatory and at the collaborating institutions in the construction and operation of the instrument.

\end{document}